\documentclass[preprint,aps,showpacs]{revtex4}
\newcommand{\sect}[1]{\setcounter{equation}{0}\section{#1}}

\begin{document}
\title{Composition law of $\kappa$-entropy for statistically independent systems}
\author{G. Kaniadakis$^1$\footnote{e-mail: giorgio.kaniadakis@polito.it}, A.M. Scarfone$^2$, A. Sparavigna$^1$, T. Wada$^3$}
\address{$^1$Department of Applied Science and Technology, Politecnico di Torino, Corso Duca degli Abruzzi 24, 10129 Torino, Italy \\ $^2$Istituto dei Sistemi Complessi (ISC-CNR) c/o Politecnico di Torino
Corso Duca degli Abruzzi 24, 10129 Torino, Italy \\ $^3$Department of Electrical and Electronic Engineering, Ibaraki University, Nakanarusawa, Hitachi, Ibaraki, 316-8511, Japan
}
\date{\today}

\begin{abstract}
The intriguing and still open question concerning the composition law of Â $\kappa$-entropy $S_{\kappa}(f)=\frac{1}{2\kappa}\sum_i (f_i^{1-\kappa}-f_i^{1+\kappa})$ with $0<\kappa<1$ and $\sum_i f_i =1$ is here reconsidered and solved. It is shown that, for a statistical system described by the probability distribution $f=\{ f_{ij}\}$, made up of two statistically independent subsystems, described through the probability distributions Â $p=\{ p_i\}$ and $q=\{ q_j\}$, respectively, with $f_{ij}=p_iq_j$, the joint entropy $S_{\kappa}(p\,q)$ can be obtained starting from the $S_{\kappa}(p)$ and $S_{\kappa}(q)$ entropies, and additionally from the entropic functionals $S_{\kappa}(p/e_{\kappa})$ and $S_{\kappa}(q/e_{\kappa})$, $e_{\kappa}$ being the $\kappa$-Napier number. The composition law of the $\kappa$-entropy is given in  closed form, and emerges as a one-parameter generalization of the ordinary additivity law of Boltzmann-Shannon entropy recovered in the $\kappa \rightarrow 0$ limit.

\end{abstract}
\pacs{05.90.+m, 05.20.-y, 51.10.+y, 03.30.+p}
\maketitle

\sect{Introduction}

The one-parameter generalized entropy defined through
\begin{eqnarray}
S_{\kappa}(f)= -\sum_i f_i \ln_{\kappa} (f_i)
\ , \ \ \ \label{1}
\end{eqnarray}
with
\begin{eqnarray}
\ln_{\kappa} (x)=\frac{x^{\kappa}-x^{-\kappa}}{2\kappa}
\ , \ \ \ \label{2}
\end{eqnarray}
and $0<\kappa<1$,  $f=\{ f_i \, ; \, Â i=1,...,N \, ; \, \sum_i f_i=1\}$, is considered to be a good candidate for the relativistic extension of Boltzmann-Shannon entropy, which is recovered in the $\kappa \rightarrow 0$ limit \cite{PRE2002,PRE2005}. The $\kappa$-logarithm $\ln_{\kappa} (x)$ emerges as the relativistic generalization of the ordinary logarithm $\ln(x)$, which is obtained in the same limit. Its inverse function, namely the $\kappa$-exponential, is given by
\begin{equation}
\exp_{\kappa}(x)=(\sqrt{1+ \kappa^2 x^2}+\kappa x)^{1/\kappa} \ \ ,
\label{3}
\end{equation}
and it is obtained within the special relativity, starting directly from the dynamical Lorentz transformations \cite{PLA2011}.

It has been shown that, in case of a particle system with energy levels $E_i$, the maximization of entropy (\ref{1}) under the $\sum_i f_i=1$ and $\sum_i f_i\, E_i=\,<\!E\!>$ constraints, according to the Maximum Entropy Principle \cite{EPJB2009}, leads to the probability distribution
\begin{equation}
f_i=\frac{1}{e_{\kappa}}\exp_{\kappa}(-\gamma_{\kappa}\beta E_i+\gamma_{\kappa}\beta \mu) \ \ , Â \label{4}
\end{equation}
where $\beta$ and $-\beta\mu$ are the usual Lagrange multipliers, while the constants $\gamma_{\kappa}$ and $e_{\kappa}$, given by
\begin{eqnarray}
&&\gamma_{\kappa}=\frac{1}{\sqrt{1-\kappa^2}} \ , \label{5} \\
&&e_{\kappa}=\left ( \frac{1+\kappa}{1-\kappa} \right )^{1/2\kappa}
\ , \ \ \ \label{6}
\end{eqnarray}
linked through $e_{\kappa}=\exp_{\kappa}(\gamma_{\kappa})$, represent a Lorentz factor and the relativistic generalization of the Napier number $e$, respectively.

The most interesting feature of the statistical distribution (\ref{4}) is undoubtedly given by its asymptotic behavior. The low energy limit of that distribution is equivalent to the $\kappa \rightarrow 0$ limit, and yields the ordinary Boltzmann-Gibbs exponential distribution
\begin{equation}
f_i \approx \exp\,(-\beta E_i+\beta \mu -1) \ \ . \label{7}
\end{equation}

On the other hand, the distribution (\ref{4}), in the high energy limit shows a power-law tail

\begin{equation}
f_i\approx \frac{1}{e_{\kappa}} \, \, (2\kappa \gamma_{\kappa} \beta)^{-1/\kappa}\, E_i^{-1/\kappa} \ \ , \label{8}
\end{equation}

which has been observed experimentally in many physical, natural and artificial systems \cite{EPJB2009}.

Generalized statistical mechanics, based on $\kappa$-entropy, preserves the main features of ordinary Boltzmann-Gibbs statistical mechanics and is related to Boltzmann relativistic nonlinear  kinetics \cite{KQSphysA2003,BiroK2006,EPL2010}. Over the last 15 years this statistical theory has attracted the interest of many researchers, who have studied its foundation and mathematical aspects \cite{Silva06A,Naudts1,Topsoe,Tempesta2011,Scarfone2013,SouzaPLA2014,Scarfone1,Scarfone2}, the underlying thermodynamics \cite{Wada1,ScarfoneWada,ScarfoneWadaJPA2014,Bento3lawThermod,WadaMatsuzoeScarfone2015}, and, at the same time, specific applications of the theory to various fields. A nonexhaustive list of applications includes, among others, those in plasma physics \cite{Lourek,Gougam,Chen}, in astrophysics \cite{Carvalho,Carvalho2,Carvalho2010,Cure,AbreuEPL,AbreuIJMPA,Chen}, in quantum statistics \cite{Santos2011a,Planck,Lourek2}, in quantum entanglement  \cite{Ourabah,OrabahPhyscripta}, in genomics \cite{SouzaEPL2014}, in complex networks \cite{Macedo,Stella}, in economy \cite{Clementi2007,Clementi2009,Bertotti,Modanese} and in finance \cite{Trivellato2012,Trivellato2013,Tapiero}.

The  $S_{\kappa}$ entropy, like Boltzmann-Shannon entropy, is thermodynamically and Lesche stable. On the other hand, while Boltzmann-Shannon entropy is additive for statistically independent systems, $\kappa$-entropy is superadditive \cite{ScarfoneWada}, but its exact composition law has still not been completely established. The present paper deals with this intriguing and still open question regarding the composition law of $S_{\kappa}$ and the our main goal has just been its derivation.

The paper is organized as follows: First, in Sec. II, it is shown that $\ln_{\kappa}(xy)$ can be decomposed into the functions $\ln_{\kappa}(x)$ and $\ln_{\kappa}(y)$. In Sec.III, by employing this result, the composition law of the joint $\kappa$-entropy of two statistically independent systems is obtained. In Sec. IV, after introducing the concept of $\kappa$-parentropy, this composition law, is re-written in a different and more elegant form, which point out better, the superadditive nature of $S_{\kappa}$. Finally, in Sec. V, the origin and physical meaning of the obtained composition law of $\kappa$-entropy, is discussed and some concluding remarks are reported.

\sect{The $\kappa$-logarithm composition law}

Let us consider the composition law of $\kappa$-logarithm \cite{PRE2002,PRE2005}
\begin{equation}
\ln_{\kappa} (xy)= Â \ln_{\kappa} (x) \sqrt{1+\kappa^2 \ln_{\kappa}^2 (y)} + \ln_{\kappa} (y) \sqrt{1+\kappa^2 \ln_{\kappa}^2 (x)}
\ , \ \ \ \label{9}
\end{equation}
which implies the important inequality
\begin{eqnarray}
\ln_{\kappa} (xy) \leq \ln_{\kappa} (x) + \ln_{\kappa} (y)
\ , \ \ \ \label{10}
\end{eqnarray}
holding in the physically meaningful interval $0<x,y\leq 1$.
It is remarkable that this composition law is, in fact just the composition law of the relativistic momenta in special relativity \cite{PRE2002,PRE2005}.

It is useful for the present discussion to note that
\begin{eqnarray}
\sqrt{1+\kappa^2 \ln_{\kappa}^2 (x)}= &&\!\!\!\!\!\frac{x^{\kappa}+ x^{-\kappa}}{2} = \frac{x^{\kappa}- x^{-\kappa}}{2\kappa} \nonumber \\ \nonumber
-&&\!\!\!\!\! \frac{(1-\kappa)x^{\kappa}- (1+\kappa)x^{-\kappa}}{2\kappa} \\
=&&\!\!\!\!\!\ln_{\kappa} (x)- \sqrt{1-\kappa^2}\,\ln_{\kappa}\!\left (\sqrt{\frac{1-\kappa}{1+\kappa}} \, \, x \right)
\ , \ \ \ \nonumber \label{}
\end{eqnarray}
so that, after taking into account the definitions of the constants $\gamma_{\kappa}$ and $e_{\kappa}$, given by Eq.s (\ref{5}) and (\ref{6}), respectively, it follows
\begin{eqnarray}
\sqrt{1+\kappa^2 \ln_{\kappa}^2 (x)}=\ln_{\kappa} (x)-\frac{1}{\gamma_{\kappa}}\,\ln_{\kappa}\left ( x/e_{\kappa}\right )
\ . \ \ \ \label{11}
\end{eqnarray}

Starting from this relation, the composition law  of $\kappa$-logarithm (\ref{9}) can be written in the following suggestive form
\begin{eqnarray}
\ln_{\kappa} (xy)=&& \!\!\!\!\! \ln_{\kappa} (x) \left ( \,\ln_{\kappa} (y) - \frac{1}{\gamma_{\kappa}} \, \ln_{\kappa} (y/e_{\kappa}) \right ) Â \nonumber \\
+ && \!\!\!\!\! \ln_{\kappa} (y) \left (\, \ln_{\kappa} (x) - \frac{1}{\gamma_{\kappa}} \, \ln_{\kappa} (x/e_{\kappa}) \right )
\ , \ \ \ \label{12}
\end{eqnarray}
which, in the $\kappa \rightarrow 0$ limit, reduces to the standard additivity law of the ordinary logarithm $\ln (xy)=\ln (x)+\ln (y)$.

\sect{The $\kappa$-entropy composition law}

Let us consider the set of normalized probability distributions
$f=\{ f_i \, ; \, i=1,...,N\}$. In ref. \cite{PRE2002}, the $\kappa$-entropy related to this set of probabilities is defined by
\begin{eqnarray}
S_{\kappa}(f)= -<\ln_{\kappa} (f_i)> \ \ . \ \ \ \label{13}
\end{eqnarray}
This entropy, which assumes the form given in Eq. (\ref{1}), after maximization according to the Maximum Entropy Principle, yields the probability distribution (\ref{4}).

Starting from the set of normalized probability distributions $f=\{f_i\}$, let us introduce the set of non-normalized distributions, defined according to $f^*=\{ f^*_i=f_i/e_{\kappa}; i=1,...,N ; \sum_i f^*_i=1/e_{\kappa}<1\}$. The $\kappa$-entropy related to that particular set of distributions is defined by
\begin{eqnarray}
S_{\kappa}(f^*)= -<\ln_{\kappa} (f^*_i)> = -\frac{\sum_i f^*_i \ln_{\kappa} (f^*_i)}{\sum_i f^*_i } \ , \ \ \ \label{14}
\end{eqnarray}
and assumes the form
\begin{eqnarray}
S_{\kappa}(f^*)= -e_{\kappa} \sum_i f^*_i \ln_{\kappa} (f^*_i)
\ . \ \ \ \label{15}
\end{eqnarray}
The latter entropy can be written explicitly in terms of distributions $f$ as follows
\begin{eqnarray}
S_{\kappa}(f/e_{\kappa}) \!=\! -\!\sum_i f_i \ln_{\kappa} (f_i/e_{\kappa})
\ . \ \ \ \label{16}
\end{eqnarray}
In the $\kappa \rightarrow 0$ limit, the functional $S_{\kappa} (f/e_{\kappa})$, reduces to $S(f/e)$ which is related to Boltzmann-Shannon entropy $S(f)$, through $S(f/e)=S(f)+1$. The functional $S(f/e)$ is regularly used in the ordinary kinetic theory in place of the Boltzmann-Shannon entropy.

It is remarkable that the maximization of the functional $S_{\kappa} (f/e_{\kappa})$, under the standard constrains, yields the distribution
\begin{equation}
f_i=\exp_{\kappa}(-\gamma_{\kappa}\beta E_i+\gamma_{\kappa}\beta \mu) \ \ , Â \label{17}
\end{equation}
instead of the distribution given in Eq. (\ref{4}).

Let us consider two statistically independent systems described by the normalized probability distributions $p=\{ p_i \, ; \, i=1,...,N \}$ and $q=\{ q_i \, ; \, i=1,...,M \}$, respectively. Furthermore, the probability distribution of the composed system is indicated by $f=p\,q$ with $f=\{ f_{ij}=p_iq_j \, ; \, i=1,...,N \,;\, j=1,...,M \, ; \, \sum_{ij} f_{ij}=1\}$. After taking into account the composition law of the $\kappa$-logarithm given in Eq. (\ref{12}), and the definitions of the entropic functionals given by Eq.s (\ref{1}) and (\ref{16}), the following composition law is obtained
\begin{eqnarray}
S_{\kappa} (p\,q)=&& \!\!\!\!\! S_{\kappa} (p) \left ( \frac{1}{\gamma_{\kappa}} \, S_{\kappa} (q/e_{\kappa}) -S_{\kappa} (q) \right ) Â \nonumber \\
+ && \!\!\!\!\! S_{\kappa} (q) \left ( \frac{1}{\gamma_{\kappa}} \, S_{\kappa} (p/e_{\kappa}) - S_{\kappa} (p) \right )
\ . \ \ \ \label{18}
\end{eqnarray}
The above composition law for the entropy of the system composed by the two statistically independent subsystems, in the $\kappa \rightarrow 0$ limit, reduces to the standard additivity law $S(p\,q)=S(p)+S(q)$ of the ordinary statistical mechanics. The subadittivity or superadittivity nature of the composition law (\ref{18}), will be the task of the next section.

It is worth mentioning that the structure of the above composition law i.e.
\begin{eqnarray}
S_{\kappa} (p\,q)= S_{\kappa} (p) \,{\cal I}_{\kappa} (q)+ S_{\kappa} (q)\, {\cal I}_{\kappa} (p)
\ . \ \ \ \label{19}
\end{eqnarray}
was first noted in ref. \cite{ScarfoneWada}, where the functional ${\cal I}_{\kappa} (p)$ was introduced without connecting it to ${S}_{\kappa} (p)$. In this sense, the missed connection, which has here been established between ${\cal I}_{\kappa} (p)$ and ${S}_{\kappa} (p)$ i.e.
\begin{eqnarray}
{\cal I}_{\kappa} (p)= \frac{1}{\gamma_{\kappa}} \, S_{\kappa} (p/e_{\kappa}) - S_{\kappa} (p)
\ , \ \ \ \label{20}
\end{eqnarray}
represents one of the main achievements of the present work.

\sect{The $\kappa$-Parentropy}

In order to obtain a better understanding of the origin and meaning of the function $\ln_{\kappa}(x/e_{\kappa})$ that appears in the composition law (\ref{12}), the concept of scaled $\kappa$-logarithm introduced in ref. \cite{EPJB2009}, is breafly recalled. The scaled $\kappa$-logarithm is a two-parameter deformed logarithm that can be constructed, starting from the $\kappa$-logarithm and the scaling parameter $\varsigma$, according to
\begin{eqnarray}
\ln_{\kappa \varsigma} (x)= \frac{\,\ln_{\kappa} (\varsigma x)-\ln_{\kappa}(\varsigma)}{\sqrt{1+\kappa^2\,\ln_{\kappa}^2(\varsigma)}}
\ . \label{21}
\end{eqnarray}

It is worth mentioning that the scaled $\kappa$-logarithm $\ln_{\kappa \varsigma} (x)$, represents a family of deformed logarithms that contains the $\ln_{\kappa} (x)$ recovered when $\varsigma=1$, as a special case. This family also contains the functions $\ln_{q} (x)$ and $\ln_{2-q} (x)$ of the nonextensive statistical mechanics, obtained in the $\varsigma \rightarrow 0^+$  and $\varsigma \rightarrow +\infty$ limits, respectively, after posing $\kappa=q-1$ \cite{EPJB2009}. In this sense, $\ln_{\kappa \varsigma} (x)$ can also be viewed as a function that interpolates between $\ln_{\kappa} (x)$ and $\ln_{q} (x)$.

Let us focus attention on a particular scaled $\kappa$-logarithm that is obtained from the general formula (\ref{21}) after fixing $\varsigma=1/e_{\kappa}$. This scaled $\kappa$-logarithm is hereafter called $\kappa$-paralogarithm, and is indicated as $\ln_{\kappa}^* (x)$. It is easy to verify that
\begin{eqnarray}
\ln_{\kappa}^* (x)= \frac{1}{\gamma_{\kappa}}\,\ln_{\kappa}\left ( \,x/e_{\kappa}\right ) + 1
\ . \ \ \ \label{22}
\end{eqnarray}

After taking into account Eq. (\ref{11}), it is possible to express $\ln_{\kappa}^* (x)$ in terms of $\ln_{\kappa} (x)$ according to
\begin{eqnarray}
\ln_{\kappa}^* (x) = \ln_{\kappa} (x) - \sqrt{1+\kappa^2 \ln_{\kappa}^2 (x)} +1
\ , \ \ \ \label{23}
\end{eqnarray}
from which it follows that $\ln_{\kappa}^* (x) \leq \ln_{\kappa} (x)$.

The $\ln_{\kappa}^* (x)$ function, defined in ${\bf R}^+$, increases monotonically i.e. $d\ln_{\kappa}^* (x)/dx > 0$ and is convex i.e. $d^2\ln_{\kappa}^* (x)/dx^2 > 0$. Furthermore, the following Taylor expansions holds
\begin{equation}
\ln_{\kappa}^*(1+x)
{\atop\stackrel{\textstyle\sim}{\scriptstyle x\rightarrow {0}}}x-
\left(\!1+\,\kappa^2\right) \frac{x^2}{2} \ , \label{24}
\end{equation}
while the power-law asymptotic behaviors of $\ln_{\kappa}^*(x)$ are given by
\begin{eqnarray}
&&\ln_{\kappa}^*(x)
{\atop\stackrel{\textstyle\sim}{\scriptstyle x\rightarrow
{0^+}}}-\frac{1+\kappa}{2\kappa} \,x^{-\kappa} \,\,
Â \ , \ \ \ \ \ \ \ \label{25} \\
&&\ln_{\kappa}^*(x)
{\atop\stackrel{\textstyle\sim}{\scriptstyle x\rightarrow
{+\infty}}}\,\frac{1-\kappa}{2\kappa}\, x^{\,\kappa} \,\, \ , \label{26}
\end{eqnarray}
and finally it holds
\begin{equation}
\int_0^1\ln_{\kappa}^*(x)dx = -\,\frac{1+\kappa^2}{1-\kappa^2}
 \ \ . \label{27}
\end{equation}

After introducing the $\kappa$-paralogarithm $\ln_{\kappa}^*(x)$, the composition law of $\ln_{\kappa}(xy)$, given in Eq. (\ref{12}), assumes the following alternative form
\begin{eqnarray}
\ln_{\kappa} (xy)=&& \!\!\!\!\! \ln_{\kappa} (x) \big [ 1+ \ln_{\kappa} (y) - \ln_{\kappa}^* (y) \big ] Â \nonumber \\
+ && \!\!\!\!\! \ln_{\kappa} (y) \big [ 1+ \ln_{\kappa} (x) - \ln_{\kappa}^* (x) \big ]
\ . \ \ \ \label{28}
\end{eqnarray}
From the latter composition law, and after taking into account that $\ln_{\kappa}^* (x) \leq \ln_{\kappa} (x)$, it follows the property $\ln_{\kappa} (xy) \leq Â \ln_{\kappa} (x) + \ln_{\kappa} (y)$, holding for $0<x,y \leq 1$.

Let us introduce the $\kappa$-parentropy of a normalized set of probability distributions $f=\{f_i\}$ defined by
\begin{eqnarray}
S_{\kappa}^*(f)= -<\ln_{\kappa}^* (f_i)> \, = -\sum_i f_i \ln_{\kappa}^* (f_i)
\ . \ \ \ \label{29}
\end{eqnarray}
After taking into account the definition of the $\kappa$-paralogarithm (\ref{22}), it is easy to verify that $S_{\kappa}^*(f)$ is related to the functional $S_{\kappa} (f/e_{\kappa})$ through the linear transformation
\begin{eqnarray}
S_{\kappa}^* (f)= \frac{1}{\gamma_{\kappa}} \, S_{\kappa} (f/e_{\kappa}) - 1
\ . \ \ \ \label{30}
\end{eqnarray}

In order to explain the meaning of the parentropy concept $S_{\kappa}^* (f)$ more clearly, let us consider a statistical system described by the ordered set of normalized probability distributions $f=\{ f_i=\delta_{i1}; 1\leq i \leq N \}$. This system occupies the state labeled by the index $i=1$, and the entropy of the system is consequently equal to zero. It is easy to verify that $S_{\kappa} (f)= S_{\kappa}^* (f)=0$, therefore entropy $S_{\kappa} (f)$ and parentropy $S_{\kappa}^* (f)$ can be adopted to describe the system. Otherwise it results that $S_{\kappa} (f/e_{\kappa})=\gamma_{\kappa} \neq 0$, and the functional $S_{\kappa} (f/e_{\kappa})$ therefore cannot represent the entropy of the system.

Finally, the composition law (\ref{18}), after taking into account the relationship (\ref{30}), assumes the following alternative form
\begin{eqnarray}
S_{\kappa} (p\,q)=&& \!\!\!\!\! S_{\kappa} (p) \big [ 1+ S_{\kappa}^* (q) - S_{\kappa} (q) \big ] Â \nonumber \\
+ && \!\!\!\!\! S_{\kappa} (q) \big [ 1+ S_{\kappa}^* (p) - S_{\kappa} (p) \big ]
\ , \ \ \ \label{31}
\end{eqnarray}
while the functional ${\cal I}_{\kappa} (p)$, defined in (\ref{20}), assumes the form
\begin{eqnarray}
{\cal I}_{\kappa} (p)= 1 + S_{\kappa}^* (p) - S_{\kappa} (p)
\ . \ \ \ \label{32}
\end{eqnarray}

After recalling that $\ln_{\kappa}^* (x) \leq \ln_{\kappa} (x)$, and then $S_{\kappa}^* (p) \geq S_{\kappa}(p)$ and ${\cal I}_{\kappa} (p)\geq 1$, it follows from (\ref{31}), that the $\kappa$-entropy is superadditive i.e. $S_{\kappa} (p\,q) \geq S_{\kappa} (p) + S_{\kappa}(q)$. Therefore, the difference between the entropy of the system obtained from the composition of two statistically independent subsystems and the sum of the entropies of the two individual subsystems i.e.
\begin{eqnarray}
\Delta S_{\kappa} = S_{\kappa} (p\,q) - S_{\kappa} (p) - S_{\kappa} (q)
\ , \ \ \ \label{33}
\end{eqnarray}
results to be $\Delta S_{\kappa}\geq 0$, and is given by
\begin{eqnarray}
\Delta S_{\kappa}=S_{\kappa} (p) Â S_{\kappa}^* (q) Â + S_{\kappa} (q) S_{\kappa}^* (p) - 2 S_{\kappa} (p) S_{\kappa} (q)
\ . \ \ \ \label{35}
\end{eqnarray}

\sect{Discussion}

The question regarding the physical meaning of the entropy excess $\Delta S_{\kappa}$, between the entropy of the composite system and the sum of the entropies of the two individual statistically independent subsystems is an important, and at the same time, very complex issue which should be considered within the context of the broader problem, which is still under intensive discussion, pertaining the origin of the generalized entropies, regardless on their particular form. It can be noted that if the classical Maxwell-Boltzmann statistics, corresponding to the additive Boltzmann-Shannon entropy is excluded, all the other statistics considered in the literature are associated with particular types of nonadditive generalized entropies.

A first remark that deserves attention regards the thermodynamic stability of system described by a generalized entropy. In the case of a system interacting with a bath in thermodynamic equilibrium, the particle density describing the state of the system in stationary conditions, is obtained starting from the maximum entropy principle. Alternatively the particle density function can be obtained as the asymptotic solution for $t \rightarrow \infty$, of a nonlinear Fokker-Planck evolution equation, obeying the second law of thermodynamics. On the other hand, this is also the case of an isolated system whose kinetics is governed by a nonlinear Boltzmann equation obeying the H-theorem that accounts the thermodynamics second law. This theorem states that the generalized entropy of the system grows continuously as it evolves towards equilibrium, where the entropy becomes maximum and reaches a finite value. The fact that $\kappa$-entropy is superadditive, and the entropy of the composite system is greater than the sum of the entropies of the individual subsystems, indicates that the composite system is thermodynamically more stable. This applies to all systems described by generalized superadditive entropies, and it was first noted by Landsberg and Vedral \cite{Landsberg}.

As far as the origin of the nonadditivity of generalized entropies is concerned,  it can be observed that the systems described by these entropies are nonlinear. It is known that Boltzmann-Shannon entropy is associated with the ordinary Fokker-Planck, or Boltzmann, equations, depending on whether the system is in interaction with a bath or is isolated. In both cases, these equations describe linear kinetics. However the generalized entropies are instead associated with non-linear kinetics, governed by particular  nonlinear evolution equations. In the case of systems subjected to $\kappa$-entropy, the corresponding  nonlinear Fokker-Planck, or nonlinear Boltzmann, equations are known in the literature \cite{PRE2002,EPL2010}.

In ref. \cite{ScarfoneWadaJPA2014}, the thermodynamic potentials i.e. the free energy and the Massieu function have been obtained within the $\kappa$-statistical mechanics, and it has been shown that $\kappa$-thermodynamics preserves the Legendre structure of ordinary thermodynamics. Therefore, the nonadditivity of the $\kappa$-entropy implies the nonadditivity of the system internal energy. More in general and independently on the form of the generalized entropy, the positive sign of $\Delta S$ indicates the existence of an attractive force between the two statistically independent subsystems, and this interaction clearly originates from the interactions between the constituent particles of any subsystem. On the other hand, the interaction between the two subsystems is only negligible in the case of short range particle interaction. For this reason, the hypothesis has been advanced in the literature, that the systems described by generalized entropies could be composed of particles with long range interactions.

Furthermore, another important question deserves to be addressed i.e. the connection between nonadditive systems and ordinary additive thermodynamics. In \cite{Ruffo}, the authors claimed that a large ensemble of replicas of a nonadditive system obeys the standard  additive  thermodynamics principles, and the formalism of ordinary thermodynamics can be naturally applied, and the properties of nonadditive systems can be inferred.

Let us now briefly discuss the origin of $\kappa$-entropy. First it should be noted that in the case Fermi-Dirac or Bose-Einstein statistics the nonlinear terms in the expression of the generalized fermionic o bosonic entropy are introduced precisely by the Pauli exclusion or inclusion principle, which accounts for the particle quantum nature. The related Fokker-Planck kinetics of fermions or bosons are nonlinear \cite{ClassicalBF}. This is also the case of the Boltzmann kinetics of fermions or bosons described by the nonlinear Uehling-Uhlenbeck nonlinear equations. Analogously, the intermediate between the fermionic and bosonic nature of some many-body quantum systems leads to anyonic statistics i.e. Haldane-Wu particles, quons etc, which are described by generalized entropies.

Finally, let us consider $\kappa$-statistics. It should also be noted that $\kappa$-entropy emerges within the Einstein's special relativity. The fundamental law of relativistic dynamics governing the time evolution of any particle of the macroscopic statistical system, i.e. the Newton relativistic equation, is a non-linear equation. As a consequence, according to the standard principles of molecular dynamics, a many-body system composed of relativistic particles  necessarily is governed by nonlinear kinetics and thermodynamics. The related nonlinear kinetic equation of this statistical system is known in the literature and it leads directly to $\kappa$-entropy \cite{PRE2002,PRE2005,PLA2011}. Furthermore, the emergence of superadditive macroscopic quantities such as $\kappa$-entropy, should not be surprising if it is taken into account that superadditive physical quantities can also be found at a microscopic level. In fact, it is well known, in relativistic dynamics, that the proper mass of an N-particle system is superadditive.

It has long been long known that the relativistic particles that form the cosmic rays violate Maxwell-Boltzmann statistics, or equivalently, Boltzmann-Shannon entropy. On the other hand, given the large extension of the cosmic ray spectrum i.e. 33 decades in particle flux and 13 decades in energy, the excellent agreement observed over so many decades between these experimental data and the theoretical predictions of $\kappa$-statistical mechanics is quite remarkable. This result represents important experimental evidence of the relativistic origin of $S_{\kappa}$ \cite{PRE2002}.

\end{document}